\documentclass[aps,prl,amsmath,amssymb,reprint,superscriptaddress,footinbib]{revtex4-1}

\usepackage{graphicx}
\usepackage{units}
\usepackage{braket}
\usepackage{hyperref}

\begin{document}
	
\preprint{}

\newcommand{\thetitle}{Quantum harmonic oscillator spectrum analyzers}

\title{\thetitle}

\author{Jonas Keller}
\affiliation{National Institute of Standards and Technology, 325 Broadway, Boulder, CO 80305, USA}
\affiliation{Department of Physics, University of Colorado, Boulder, CO 80309, USA}
\affiliation{Present Address: Physikalisch-Technische Bundesanstalt, Bundesallee 100, 38116 Braunschweig, Germany}
\author{Pan-Yu Hou}
\affiliation{National Institute of Standards and Technology, 325 Broadway, Boulder, CO 80305, USA}
\affiliation{Department of Physics, University of Colorado, Boulder, CO 80309, USA}
\author{Katherine C.~McCormick}
\affiliation{National Institute of Standards and Technology, 325 Broadway, Boulder, CO 80305, USA}
\affiliation{Department of Physics, University of Colorado, Boulder, CO 80309, USA}
\affiliation{Present Address: Department of Physics, University of Washington, Seattle, WA 98195, USA}
\author{Daniel C.~Cole}
\affiliation{National Institute of Standards and Technology, 325 Broadway, Boulder, CO 80305, USA}
\author{Stephen D.~Erickson}
\affiliation{National Institute of Standards and Technology, 325 Broadway, Boulder, CO 80305, USA}
\affiliation{Department of Physics, University of Colorado, Boulder, CO 80309, USA}
\author{Jenny J.~Wu}
\affiliation{National Institute of Standards and Technology, 325 Broadway, Boulder, CO 80305, USA}
\affiliation{Department of Physics, University of Colorado, Boulder, CO 80309, USA}
\author{Andrew C.~Wilson}
\affiliation{National Institute of Standards and Technology, 325 Broadway, Boulder, CO 80305, USA}
\author{Dietrich Leibfried}
\affiliation{National Institute of Standards and Technology, 325 Broadway, Boulder, CO 80305, USA}
\date{\today}

\begin{abstract}
Characterization and suppression of noise are essential for the control of harmonic oscillators in the quantum regime. We measure the noise spectrum of a quantum harmonic oscillator from low frequency to near the oscillator resonance by sensing its response to amplitude modulated periodic drives with a qubit. Using the motion of a trapped ion, we experimentally demonstrate two different implementations with combined sensitivity to noise from $\unit[500]{Hz}$ to $\unit[600]{kHz}$. We apply our method to measure the intrinsic noise spectrum of an ion trap potential in a previously unaccessed frequency range.
\end{abstract}

\maketitle

Harmonic oscillators (HOs) are ubiquitous in physics, describing such diverse phenomena as molecular vibrations, the baryon acoustic oscillations in the early universe \cite{Eisenstein2005}, electromagnetic fields, and normal or superconducting electrical circuits. Some HO systems---for example micro-resonators \cite{Aspelmayer2014}, the motion of neutral atoms \cite{Grimm00} and ions \cite{Leibfried2003} in trapping potentials, photons in optical and microwave resonators \cite{Haroche2013}, and vibrations in solids \cite{Lee2011, Hou2016}---can be controlled in the quantum regime. Precisely controlled HO systems feature prominently in precision metrology \cite{LIGO13}, fundamental quantum mechanical research \cite{Haroche2013}, and quantum information processing (QIP) \cite{Ladd10}. QIP uses quantum-controlled HO degrees of freedom either as units of quantum information in their own right \cite{Gottesman2001,Fluehmann2019,Campagne-Ibarcq2020} or as a mechanism to couple identical \cite{Cirac1995,Blais04,Aspelmayer2014} or distinct quantum bits (qubits) \cite{Xiang2013,Andrews2014,Kurizki2015,Kotler2017}. In all instances, noise limits the practical coherence of HOs, which makes proper noise characterization desirable.\\

For two-level systems, techniques pioneered in the field of nuclear magnetic resonance (such as the Hahn echo \cite{Hahn1950} and Carr-Purcell-Meiboom-Gill (CPMG) sequences \cite{Carr1954,Meiboom1958}) as well as dynamical decoupling \cite{Viola1999} and adaptations \cite{Uhrig2007,Biercuk2009a}, are routinely employed to suppress sensitivity to noise at certain frequencies or to characterize its spectrum \cite{Lasic2006,Alvarez2011,Bylander2011,Kotler2013,Bishof2013}. Here, we apply similar principles to an HO and make use of its larger Hilbert space to simplify spectrum reconstruction and remove ambiguities. We demonstrate two different methods experimentally with a trapped ion and explore previously unaccessed regions of its motional spectrum. Manipulation and readout of the HO state is achieved in our case via a Jaynes-Cummings (JC) type coupling to a qubit \cite{Blockley1992}. This concept can be applied in other HO systems as well, and experimental capabilities suitable for our techniques have been demonstrated, e.g. with microwave \cite{Hofheinz2009} and acoustic \cite{Chu2018} resonators.\\

We would like to link fluctuations $\Delta(t)$ of the HO's angular frequency over time $t$, which we assume to be ergodic and stationary, to a time-independent, two-sided power-spectral density (PSD) $S_\Delta(f)$. For infinite time resolution and sampling duration, $S_\Delta(f)$ can be obtained from a time series record of $\Delta(t)=\omega(t)-\omega_0$, referenced to a local oscillator (LO) at $\omega_0$, as the Fourier transform of the autocorrelation function of $\Delta(t)$. In practical frequency analysis with finite time resolution and sampling duration, a filter $\tilde{s}(f)$ typically removes frequency contributions to $S_\Delta(f)$ outside a band of interest around a center frequency $f_0$. Popular choices are Gaussian or Blackman \cite{Blackman1958b} filters of a certain width $\delta f$. The chosen filter can be related to a sensitivity function $s(t)$ applied to $\Delta(t)$ by \cite{Blackman1958a,Supplemental} 
\begin{equation}
\label{PSD_result}
\langle\phi^2\rangle:=\int_{-\infty}^\infty \left\vert \tilde{s}(f)\right\vert^2S_\Delta(f)df=
\left\langle\left\vert\int_{-t_w}^{t_w}s(t)\Delta(t)dt
\right\vert^2\right\rangle\;\textnormal{,}
\end{equation}
where $\langle\phi^2\rangle$ is proportional to the spectral power inside the filter, $s(t)$ is the Fourier transform of $\tilde{s}(f)$ and the measurement duration $2~t_w$ has to be chosen such that $s(t)$ approximately vanishes outside $[-t_w,t_w]$. The physical meaning of the quantity $\phi$ depends on the implementation.
For a resolution bandwidth of $\delta_\mathrm{rbw}$, defined here as the full-width at half-maximum (FWHM) of $\vert\tilde{s}(f)\vert^2$, we define the amplification $a_{\tilde{s}}$ as
\begin{equation}
\label{amplification_definition}
a_{\tilde{s}}:=\left(\int_{-\infty}^\infty\vert \tilde{s}(f)\vert^2df\right)/\delta_\mathrm{rbw}\;\textnormal{,}
\end{equation}
and approximate the PSD around $f_0$, filtered by $\tilde{s}(f)$, by
\begin{equation}
\label{PSD_approximation}
\tilde{S}_\Delta(f_0) = \langle\phi^2\rangle/(a_{\tilde{s}}~ \delta_\mathrm{rbw})\;\textnormal{.}
\end{equation}
For $\delta_{\rm rbw}\rightarrow 0$, $\tilde{S}_\Delta(f)$ approaches $S_\Delta(f)$.\\

Smoothly varying envelopes $s(t)$ can suppress side lobes and harmonics in $\vert \tilde{s}(f)\vert^2$ and thus simplify the interpretation of $\tilde{S}_\Delta(f)$ and the approximate reconstruction of $S_\Delta(f)$. We apply this principle to two techniques we have previously demonstrated with square-wave filters: $s(t)$ is either implemented by coherently driving the HO \cite{McCormick2019QST}, or it is approximated by a function with discrete steps for which we prepare different number states of the HO \cite{McCormick2019nature}. While the discrete approximation requires more experimental control---in our case provided by the JC coupling to a qubit---it can realize the same amplification $a_{\tilde{s}}$ with states of lower average energy compared to the realization based on coherent driving \cite{McCormick2019nature}. The two other capabilities required by both methods are enabled by JC coupling as well: Initialization of the HO in its ground state, and a second-order readout process $f(\phi)\propto\phi^2$ that determines the noise power via the second moment $\langle \phi^2\rangle\propto\langle f(\phi)\rangle$ of the respective linear HO response $\phi$.\\

\begin{figure}
	\centerline{\includegraphics[width=.5\textwidth]{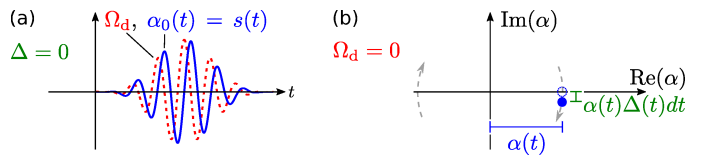}}
	\caption{\label{tickle_concept_figure}
	Implementation of filter functions $s(t)$ using coherent displacements. (a) In the absence of noise ($\Delta(t)=0$), the time-dependent displacement $\alpha_0(t)$ along the real axis (solid line) is proportional to the integral over time of the coherent drive $\Omega_d(t)$ (dashed line). (b) In the LO reference frame rotating at $\omega_0$, frequency fluctuations $\Delta(t)$ cause azimuthal rotations. During an infinitesimal time step $dt$, this produces a displacement perpendicular to $\alpha(t)$ and equal to $-i \alpha(t) \Delta(t) dt$. For small rotations, $\mathrm{Re}[\alpha(t)]\approx\alpha_0(t)$ thus controls the sensitivity to $\Delta(t)$ and implements a filter function $s(t)$.}
\end{figure}

Coherent displacements of the HO state are frequently implemented with a resonant force due to a classical field \cite{Carruthers1965}. A force resonant with the LO frequency $\omega_0$ results in the Hamiltonian
\begin{equation}\label{coherent_drive_hamiltonian}
H_\mathrm{d}(t)=-2\hbar\Omega_\mathrm{d}(t)\left(a+a^\dagger\right)\sin(\omega_0t)\;\textnormal{.}
\end{equation}
We use the time-dependent drive amplitude $\Omega_\mathrm{d}(t)$ to define the filter function $s(t)$. When initialized in the ground state, the HO remains in a coherent state $\ket{\alpha(t)}$ under the influence of the drive Eq.~(\ref{coherent_drive_hamiltonian}) and fluctuations $\Delta(t)$. In the LO reference frame, $\alpha(t)$ obeys the equation of motion \cite{Supplemental}
\begin{equation}\label{coherent_state_eom}
\dot{\alpha}(t)= \Omega_\mathrm{d}(t)-i\alpha(t)\Delta(t)\;\textnormal{.}
\end{equation}
Integrating Eq.~(\ref{coherent_state_eom}) in the absence of noise ($\Delta(t)=0$) leads to a trajectory $\alpha_0(t)$ that takes real values at all times. With a drive amplitude $\Omega_\mathrm{d}(t)=\dot{s}(t)$, the boundary conditions $\alpha_0(-t_w) =\alpha_0(t_w) \simeq 0$ hold and the displacement $\alpha_0(t)=s(t)$ can act as a sensitivity function. Figure \ref{tickle_concept_figure} (a) shows  an example for $\Omega_\mathrm{d}(t)$ and its associated displacement pattern $\alpha_0(t)$, which is a sinusoidal oscillation inside a Blackman envelope.\\

Fluctuations of the HO frequency produce additional displacements. As illustrated in Fig.~\ref{tickle_concept_figure} (b), the noisy HO rotates relative to the LO with angular velocity $\Delta(t)$, which leads to displacements at a rate $-i \alpha(t) \Delta(t)$, proportional to the distance from the origin and perpendicular to $\alpha(t)$.
For small angles, $\vert\int_{-t_w}^t\Delta(\tau)d\tau\vert\ll1\;\forall\;t\in [-t_w,t_w]$, the rotational character of these displacements can be neglected, i.e. $i \alpha(t) \Delta(t) dt \approx i\alpha_0(t) \Delta(t) dt$. In this approximation, the total displacement at the end of the sequence is \cite{Supplemental}
\begin{equation}
\alpha(t_w)\approx-i\int_{-t_w}^{t_w}\alpha_0(t)\Delta(t)dt\;\textnormal{,}
\end{equation}
such that $\langle\vert\alpha(t_w)\vert^2\rangle =: \langle\phi^2\rangle$ is proportional to the noise power within $\tilde{s}(f)$ according to (\ref{PSD_result}). If the small-angle approximation is violated, higher-order terms in the noise power modify the spectral sensitivity \cite{Supplemental}. We detect $\vert\alpha(t_w)\vert^2$ via the probability of a spin flip when driving a motion-subtracting sideband of the coupled qubit-HO system \cite{Wineland1998,Home2011,McCormick2019QST,Supplemental}. This method relies on the fact that the transition is forbidden when the HO is in its ground state, and the spin flip probability is proportional to $\vert\alpha\vert^2$ for small displacements (to within $\unit[5]{\%}$ as long as $\vert\alpha\vert < 0.47$ for the experimental parameters used below \cite{Supplemental}).\\ 

\begin{figure}
	\centerline{\includegraphics[width=.5\textwidth]{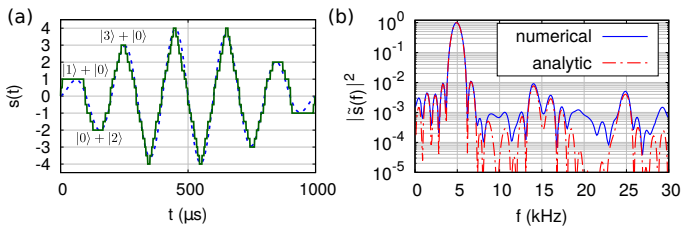}}
	\caption{\label{Fock_filter_theory}
	Implementation of filter functions $s(t)$ using number state superpositions. (a) Sensitivity function approximating a Hann filter (dashed line) by a piecewise constant sensitivity $s(t)=n_1-n_2$ (solid line) in different superpositions of number states $\ket{n_1}$ and $\ket{n_2}$, as indicated by the labels (cf.~Eq.~(\ref{Fock_superposition_state}); phase and normalization omitted for clarity). (b) The filter $\vert\tilde{s}(f)\vert^2$ corresponding to $s(t)$ shown in (a), calculated both analytically using a simple model and numerically \cite{Supplemental}.}
\end{figure}

Alternatively, the sensitivity function can be approximated in discrete steps by preparing superpositions of number states~\cite{McCormick2019nature}
\begin{equation}
  \label{Fock_superposition_state}
  \ket{\Psi_{n_1,n_2}(t)}=\frac{1}{\sqrt{2}}\left(\ket{n_1}+e^{i\phi(t)}\ket{n_2}\right)\;\textnormal{,}
\end{equation}
where $\ket{n_i}$ is the number state with $n_i$ phonons. In the interaction picture rotating with the LO, the instantaneous energy shifts of the two components are $E_i(t)=\hbar n_i\Delta(t)$, which leads to a time dependence of the relative phase described by
\begin{equation}
  \phi(t)=\phi(t_0)+\int_{t_0}^t\left(n_1-n_2\right)\Delta(\tau)d\tau\;\textnormal{.}
\end{equation}
The rate at which $\phi$ accumulates is proportional to $\Delta(t)$ and scaled by $\delta n = (n_1-n_2)$. By incrementing and decrementing $n_i$, we can produce a sensitivity function $s(t):=\delta n(t)$ with discrete steps as illustrated in Fig.~\ref{Fock_filter_theory} (a) for the example of a sinusoid within a Hann window \cite{Blackman1958a}. Figure \ref{Fock_filter_theory} (b) shows the respective frequency domain filter function $\tilde{s}(f)$. Manipulation of the motional state can be achieved via sideband transitions in the qubit-HO system \cite{McCormick2019nature,McCormick2019thesis,Supplemental}: Analogous to a Ramsey sequence, a $\pi/2$ sideband pulse transfers the system from its initial pure state into a superposition. Subsequent increments and decrements of $\delta n$ shape $s(t)$. To prepare superposition components beyond $n_i=2$, our scheme requires a three-level system in place of the qubit. At the end of the sequence, a $\pi/2$ sideband pulse maps the relative phase $\phi$ onto the qubit state populations for readout.\\

We demonstrate both methods experimentally with the axial harmonic motion of a single ${}^9$Be${}^+$ ion in a linear Paul trap. The coherent displacement experiments use the room-temperature, wafer-based, 3D trap described in \cite{Blakestad2009}, while the number state superposition method is demonstrated in a cryogenic surface-electrode trap (see \cite{Brown2011}). The distances between the ion and the nearest electrode are $\unit[160]{\mu m}$ and $\unit[40]{\mu m}$, and the axial motion is heated at a rate of ca.~$160$ and $20$ phonons per second under the respective operating conditions.\\

\paragraph{Implementation with coherent displacements.}
At the start of each experiment, the axial secular motion of the ion at $\omega_0\approx2\pi\times\unit[3.5]{MHz}$ is initialized in its ground state via resolved sideband cooling. The voltage on a nearby electrode is modulated to produce an electric field that realizes the coherent drive of Eq.~(\ref{coherent_drive_hamiltonian}). This modulation consists of a carrier signal at the LO frequency $\omega_0$ with an amplitude modulation $\propto\Omega_d(t)$, determined by the derivative of the desired sensitivity function $\alpha_0(t)=s(t)$. Our choice of a sinusoid under a Blackman envelope, with total duration $2~t_w$ and $k$ oscillations within $t_w$, results in a filter function $\tilde{s}(f)$ centered around $f_0=k/t_w$, with bandwidth $\delta_{\rm rbw} \approx 0.822/t_{\omega}$ \cite{Supplemental}. The final motional state is mapped onto internal states as described above and detected via state-dependent fluorescence. For the purpose of filter characterization, we use another DC electrode to apply an oscillating electric potential curvature as synthetic noise,
\begin{equation}
\label{sinusoidal_test_noise}
\Delta_\mathrm{test}(t)=\Delta_0\cos(2\pi f_\mathrm{noise}t+\varphi)\;\textnormal{,}
\end{equation}
with a constant amplitude $\Delta_0$ and a random phase $\varphi$ with respect to $s(t)$.

\begin{figure*}
	\centerline{\includegraphics[width=\textwidth]{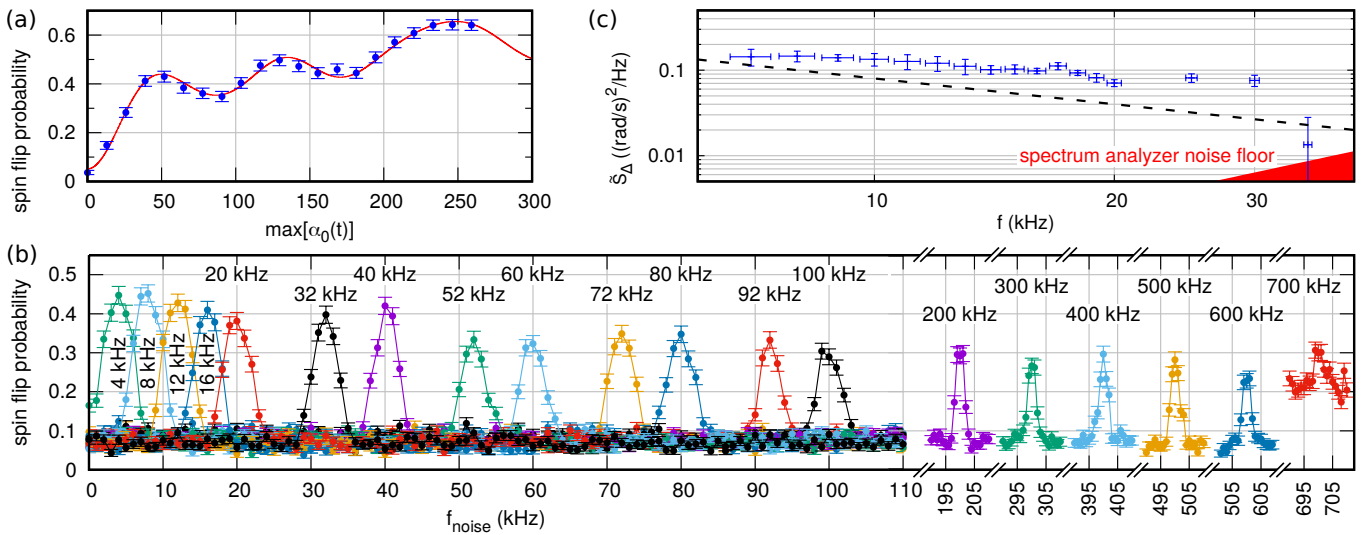}}
	\caption{\label{coherent_displacement_experiment_figure}Experimental results of the coherent displacement method. (a) Response to an $f_\mathrm{noise}=f_0=\unit[8]{kHz}$ sinusoidal modulation of the HO frequency as a function of the filter amplification (symbols: experimental data, line: model with the modulation amplitude $\Delta_0$ as a free parameter). (b) Measured response of various filter functions ($t_w=\unit[250]{\mu s}$, varied $k$) to sinusoidal modulations of varied frequency and random phase. The modulation amplitude $\Delta_0$ and filter amplification are chosen to keep the signal within the first monotonic increase of the function shown in (a). Filter functions above $\unit[100]{kHz}$ are measured with increased modulation to compensate technical limitations of the filter amplitude in this regime. Data points are linked with solid lines to guide the eye. (c) Power spectral density estimate of the HO frequency noise, determined from the transmission through filter functions with $t_w=\unit[1]{ms}$ and varied $k$. The dashed line indicates a $1/f$ slope to guide the eye. Experimental data in all subfigures are averages over $200$ repetitions. Error bars: $1\sigma$ (vertical), $\delta_\mathrm{rbw}$ (horizontal in (c))}
\end{figure*}

Figure \ref{coherent_displacement_experiment_figure} (a) shows the response to a modulation $\Delta_\mathrm{test}(t)$ with $\Delta_0=2\pi\times \unit[55]{Hz}$ at the center frequency of the filter, $f_\mathrm{noise}=\unit[8]{kHz}=k/t_w$ (where $k=2$ and $t_w=\unit[250]{\mu s}$), as a function of the filter amplitude. $\Delta_0$ is determined from the fit of a theoretical model (solid line), which includes a separately determined finite thermal energy of $0.055\,\hbar\omega_0$ due to imperfect state preparation and heating during the sequence. The initial monotonic range of this response can be used to measure the spectral sensitivity of a filter function by varying the modulation frequency $f_\mathrm{noise}$, as demonstrated in Fig.~\ref{coherent_displacement_experiment_figure} (b). We implement filter functions with $t_w=\unit[250]{\mu s}$ and different values of $k$, centered around $k\times\unit[4]{kHz}$. For data up to $f_0=\unit[100]{kHz}$, the filter and modulation amplitudes are nominally kept constant at $\Delta_0\approx2\pi\times\unit[300]{kHz}$ and $\mathrm{max}[\alpha_0(t)]=10$ throughout the measurement. We attribute signal decrease with increasing filter frequencies to a reduction of $\Delta_0$ due to imperfect compensation of the lowpass filter through which the modulation is applied. The drive $\Omega_\mathrm{d}$ required to achieve a given filter amplification $a_{\tilde{s}}\propto\mathrm{max}[\alpha_0(t)]$ increases linearly with $f_0$. We apply a rather conservative limit to the drive voltage due to unknown damage thresholds of integrated filter components, resulting in $\vert\Omega_\mathrm{d}\vert\leq\unit[2\pi\times2.1]{MHz}$, which corresponds to an electric field of $E_\mathrm{d}\leq \unitfrac[1.37]{V}{m}$ at the ion position. To observe filter functions at center frequencies above $f_0=\unit[100]{kHz}$, we instead increase the modulation amplitude $\Delta_0$. These data demonstrate the implementation of filter functions up to $f_0=\unit[600]{kHz}$ ($t_w=\unit[250]{\mu s}$, $k=150$). For $f_0=\unit[700]{kHz}$, we observe a significant increase of the background, i.e.~a displacement $\alpha(t_w)$ that is independent of $\Delta(t)$, likely due to an increased sensitivity to deviations of $\Omega_\mathrm{d}(t)$ from its ideal shape.\\

Figure \ref{coherent_displacement_experiment_figure} (c) shows an experimentally determined estimate of the trap frequency noise PSD using Eq.~(\ref{PSD_approximation}) and filter functions with $t_w=\unit[1]{ms}$ and $k=7\ldots35$ (see \cite{Supplemental} for evaluation details). It might be interpreted as localized features, e.g.~around $\unit[30]{kHz}$, on top of a $1/f$ trend (as indicated by the dashed line), as often found in technical noise sources and some models of electric field noise near surfaces \cite{Brownnutt2015}. Distinct features might be caused by specific devices in the experimental setup. In this case the spectrum analysis could aid in suppressing such contributions and further measurements focused on these regions can document improvements. In the following, we assess the frequency span and dynamic range of our implementation. The upper bound of the dynamic range is set by the linearity condition for the noise amplitude ($\vert\int_{-t_w}^t\Delta(\tau)d\tau\vert\ll1$, see above). The lowest detectable noise power is determined by shot noise and the obtainable filter amplification, which decreases linearly with $f_0$ for limited $\Omega_\mathrm{d}$. In the above example with $t_w=\unit[1]{ms}$, it is $\tilde{S}_{\Delta,\mathrm{min}}\approx7.1\times10^{-12}\unitfrac{(rad/s)^2}{Hz^3}\times f_0^2$ \cite{Supplemental}, as indicated by the shaded area in Fig.\ref{coherent_displacement_experiment_figure} (c). Finite temperature, e.g.~due to imperfect initialization or heating during the sequence, reduces detection contrast and thus increases the noise floor \cite{Supplemental}. If necessary, this limit can be reduced by averaging over more repetitions or higher amplification. The lower end of the frequency span is close to the minimum resolution bandwidth, which is determined by the maximum sequence duration with acceptable heating of the ion motion. The highest filter function frequency is limited by amplitude noise and distortion of fast drive waveforms $\Omega_\mathrm{d}$. Since the evolution according to (\ref{coherent_state_eom}) displaces the wavefunction without deformation, the readout signal could be increased by choosing a different initial state. Using number states with $n\geq1$ in place of the ground state would result in a quantum-enhanced displacement sensitivity \cite{Supplemental,Wolf2019,McCormick2019thesis}.\\

\paragraph{Implementation with number state superpositions.}
Motional state superpositions are generated using optical Raman and microwave transitions between states within the ${}^2\mathrm{S}_{1/2}$ manifold \cite{Supplemental, McCormick2019nature}. The phase $\phi$ is read out by mapping the superposition phase onto the qubit state populations, followed by a projective measurement using fluorescence detection. The resulting signal,
\begin{equation}
  \label{Fock_readout_eqn}
  P_\mathrm{bright}=\sin^2\left(\frac{\phi}{2}\right)=\frac{\phi^2}{4}+\mathcal{O}\left(\phi^4\right)\;\textnormal{,}
\end{equation}
provides the required $\phi^2$ dependence of a noise power measurement if a sufficiently small overall phase $\vert\phi\vert\ll\pi/2$ is obtained at the end of the sequence.

\begin{figure}
	\centerline{\includegraphics[width=.5\textwidth]{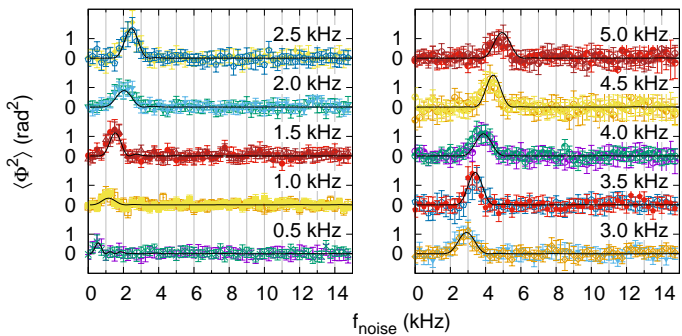}}
	\caption{\label{Fock_interferometer_figure}Number state implementation results: Measured response of different filter functions to applied modulation of varied frequency $f_\mathrm{noise}$ and random phase. The $\unit[5.0]{kHz}$ trace corresponds to the sensitivity function in Fig.~\ref{Fock_filter_theory}. The solid lines show the expected response, adjusted by a global scaling factor to account for the dependence of the modulation depth on the applied voltage. Each data point represents 200 repetitions, error bars: $1\sigma$.}
\end{figure}

We again observe the spectral sensitivities of the filter functions via sinusoidal modulation of the HO frequency, cf.~Eq.~(\ref{sinusoidal_test_noise}). The result is shown in Fig.~\ref{Fock_interferometer_figure} for ten different filter functions between $\unit[500]{Hz}$ and $\unit[5]{kHz}$. The upper limit of this frequency range, as well as the maximum sensitivity $\vert\delta n\vert$ for a given filter frequency, are determined by the pulse times needed to switch between superpositions, i.e.~by the achievable Raman Rabi frequencies. The minimum frequency and bandwidth decrease with longer sequence durations. In the presence of heating, this presents a similar trade-off with the signal-to-noise ratio as for the coherent displacement method.\\

\paragraph{Summary and conclusions.}
We have introduced a technique to measure the spectral composition of frequency fluctuations in harmonic oscillators in the quantum regime. Using a single trapped ion, we have demonstrated two different implementations. The first is based on coherent driving with a resonant force and thus simple to apply. Its signals consist of a phase-space displacement proportional to the noise power within the filter, which we measure via the coupling to a two-level system. The method may also be applicable if the HO is naturally found in the ground state and the average occupation can be read-out by other means than a sideband interaction. We have shown the coverage of a span from $\unit[4]{kHz}$ to $\unit[600]{kHz}$ with filter functions of $\unit[4]{kHz}$ resolution bandwidth using externally applied modulations as test signals. A measurement of the electric potential noise in our ion trap was performed between $\unit[7]{kHz}$ and $\unit[35]{kHz}$, limited by a conservative amplification restriction due to unknown component damage thresholds.
The second method uses a sequence of different number state superpositions that acquire a phase difference in the presence of noise, which we read out by mapping it onto a qubit. While this method requires more advanced control over the motional state, its sensitivity scales more favorably---linearly as opposed to $\propto\sqrt{E_\mathrm{max}}$---with the maximum energy $E_\mathrm{max}$ of the HO. We have generated such filter functions ranging from $\unit[500]{Hz}$ to $\unit[5]{kHz}$, with the maximum frequency limited by the minimal duration of our number state manipulations. 

These methods can be applied to any quantum harmonic oscillator for which the respective experimental capabilities for motional state manipulation and readout exist. In trapped ions, they extend the spectral range over which fluctuations of the motional frequency can be measured. This can enable better understanding and improvement of two-qubit gate fidelities \cite{Talukdar2016, Milne2020} and provide new insights into electric field noise from nearby electrode surfaces and the resulting anomalous heating \cite{Brownnutt2015}.

A complementary approach using Schr\"odinger cat states has recently been demonstrated in \cite{Milne2021}.\\

\begin{acknowledgments}
We thank D.~R.~Leibrandt and J.~F.~Niedermeyer for a careful reading of the manuscript, and D.~H.~Slichter for helpful advice and discussions. At the time the work was performed, J.K., P.-Y.H., K.C.M., S.D.E., and J.J.W.~were Associates in the Professional Research Experience Program (PREP) operated jointly by NIST and the University of Colorado under award 70NANB18H006 from the U.S. Department of Commerce, National Institute of Standards and Technology. This work was supported by the NIST Quantum Information Program and IARPA. J.K.~acknowledges support by the Alexander von Humboldt foundation. K.C.M.~acknowledges support by an ARO QuaCGR fellowship through grant W911NF-14-1-0079. D.C.C.~is supported by a National Research Council postdoctoral fellowship. S.D.E.~acknowledges support from the National Science Foundation under grant DGE 1650115. Numerical calculations were performed using QuTiP \cite{Johansson2013}. The experiments were performed using the ARTIQ control system. This Letter is a contribution of NIST, not subject to U.S. copyright.
\end{acknowledgments}
\bibliography{spectrum_analyzer_paper.bbl}
\clearpage
\pagebreak
\onecolumngrid
\begin{center}
	\textbf{\large Supplemental Material: \thetitle}
\end{center}
\setcounter{equation}{0}
\setcounter{figure}{0}
\setcounter{table}{0}
\setcounter{page}{1}
\makeatletter
\renewcommand{\theequation}{S\arabic{equation}}
\renewcommand{\thefigure}{S\arabic{figure}}

\section{Proof of Eq.~(\ref{PSD_result})}
\begin{figure}[h]
	\centerline{\includegraphics[width=.49\textwidth]{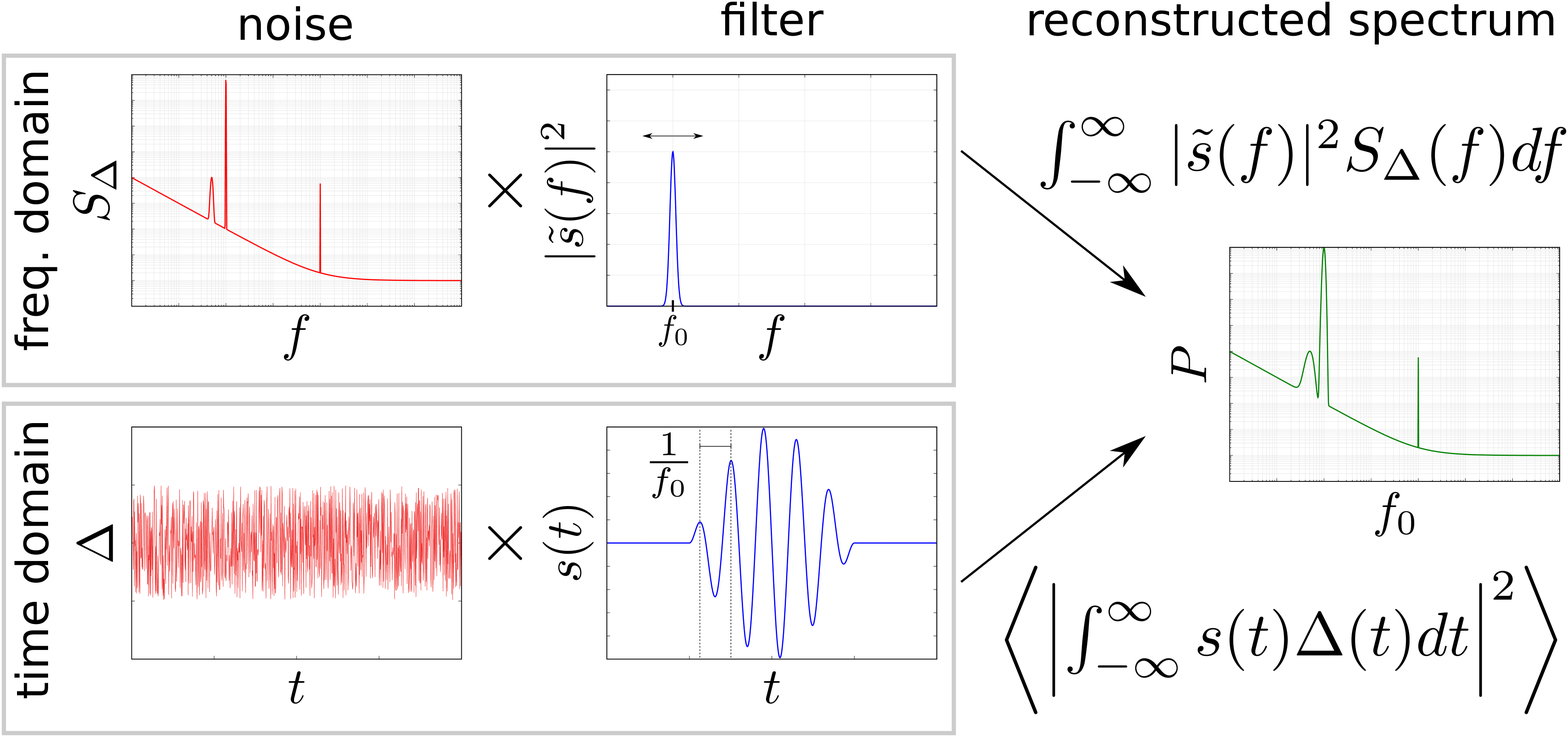}}
	\caption{\label{figure_spectrum_analyzer} Spectrum analyzer principle based on Eq.~(\ref{PSD_result}): The noise power transmitted through filter functions $\tilde{s}(f)$ with different center frequencies $f_0$ (top left) is determined as the average over many realizations of the square modulus of the time integral over the product $s(t) \Delta(t)$ (bottom left). Repeating this procedure for filters with different center frequency $f_0$ allows the reconstruction of the PSD $S_\Delta(f)$ at a resolution determined by
	$\tilde{s}(f)$.}
\end{figure}
Equation (\ref{PSD_result}) is the underlying principle of our spectrum analyzer implementations, as illustrated by Fig.~\ref{figure_spectrum_analyzer}. It can be derived as follows:
Using the fact that the filter function $s(t)$ vanishes outside the interval $[-t_w,t_w]$, the observed quantity $\langle\phi^2\rangle$ can be re-expressed as:
\begin{equation}
\left\langle\phi^2\right\rangle=\left\langle\left\vert\int_{-t_w}^{t_w} dts(t)\Delta(t)\right\vert^2\right\rangle
=\left\langle\left\vert\int_{-\infty}^\infty dts(t)\Delta(t)\right\vert^2\right\rangle\\
=\left\langle\int_{-\infty}^\infty dt\int_{-\infty}^\infty dt^\prime s(t)s^*(t^\prime)\Delta(t)\Delta^*(t^\prime)\right\rangle\;\textnormal{.}
\end{equation}
Assuming ergodicity of $\Delta$, the average over many finite-time records denoted by $\langle ... \rangle$ can be replaced by the average over a time shift $\tau$ between noise and filter in the limit of averaging over infinitely many such records:
\begin{equation}
\left\langle\phi^2\right\rangle=\lim_{T\rightarrow\infty}\frac{1}{2T}\int_{-T}^Td\tau\int_{-\infty}^\infty dt\int_{-\infty}^\infty dt^\prime s(t)s^*(t^\prime)\Delta(t+\tau)\Delta^*(t^\prime+\tau)\;\textnormal{.}
\end{equation}
Regrouping terms and making use of the time translation invariance of averages of $\Delta$ (stationarity) yields
\begin{equation}
\left\langle\phi^2\right\rangle=\int_{-\infty}^\infty dt\int_{-\infty}^\infty dt^\prime s(t)s^*(t^\prime)\underbrace{\lim_{T\rightarrow\infty}\frac{1}{2T}\int_{-T}^Td\tau\Delta(\tau)\Delta^*(\tau+t^\prime-t)}_{R_\Delta(t^\prime-t)}\;\textnormal{,}
\end{equation}
where $R_\Delta$ is the autocorrelation function of $\Delta$. We now express $s(t)$ as the Fourier transform of its frequency domain representation $\tilde{s}(f)$ and invoke the Wiener-Khinchin theorem to replace the autocorrelation function $R_\Delta$ with the Fourier transform of the power spectral density $S_\Delta$:
\begin{equation}
\left\langle\phi^2\right\rangle=\int_{-\infty}^\infty dt\int_{-\infty}^\infty dt^\prime
\left[\int_{-\infty}^\infty df^\prime \tilde{s}(f^\prime)e^{i2\pi f^\prime t}\right]
\left[\int_{-\infty}^\infty df^{\prime\prime} \tilde{s}^*(f^{\prime\prime})e^{-i2\pi f^{\prime\prime}t^\prime}\right]
\left[\int_{-\infty}^\infty df S_\Delta(f)e^{i2\pi f(t^\prime-t)}\right]\;\textnormal{.}
\end{equation}
Changing the order of integration and rearranging the arguments of the exponentials leads to Eq.~(\ref{PSD_result}) of the main text:
\begin{align}
	\left\langle\phi^2\right\rangle&=\int_{-\infty}^\infty df \int_{-\infty}^\infty df^\prime \int_{-\infty}^\infty df^{\prime\prime} \tilde{s}(f^\prime)\tilde{s}^*(f^{\prime\prime})S_\Delta(f)
	\underbrace{\int_{-\infty}^\infty dt e^{i2\pi (f^\prime-f)t}}_{\delta(f^\prime-f)}\underbrace{\int_{-\infty}^\infty dt^\prime e^{i2\pi (f-f^{\prime\prime})t^\prime}}_{\delta(f-f^{\prime\prime})}\nonumber\\
	&=\int_{-\infty}^\infty df\vert \tilde{s}(f)\vert^2S_\Delta(f)\;\textnormal{.}
\end{align}

\section{Coherent state implementation details}

\subsection{Blackman pulses}
\begin{figure}[h]
	\centerline{\includegraphics[width=\textwidth]{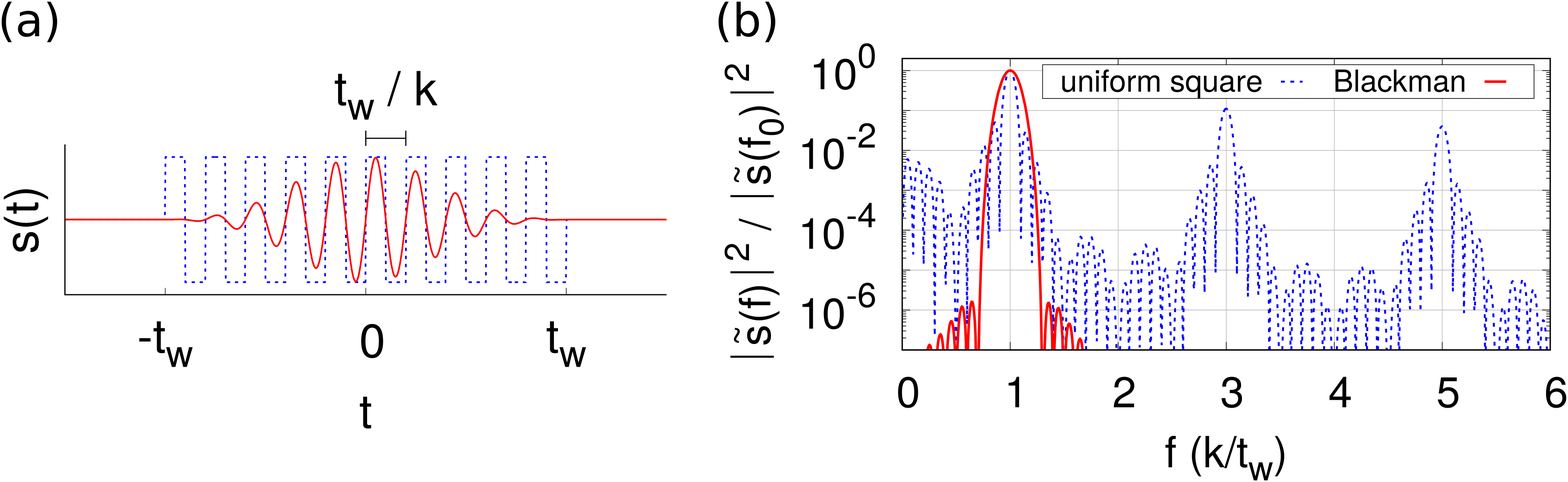}}
	\caption{\label{Blackman_vs_CPMG} Filter function consisting of a sinusoidal oscillation within a Blackman envelope. A uniform square filter function is shown for contrast to illustrate the suppression of sidelobes and harmonics by the smooth variation. (a) Time-domain filter functions. (b) Frequency domain filter functions.}
\end{figure}
For the coherent state implementation, we choose a sinusoid under a Blackman envelope \cite{Blackman1958b} of duration $2t_w$ as the time domain filter function $s(t)$ (see Fig.~\ref{Blackman_vs_CPMG} (a)):
\begin{align}
	\label{blackman_envelope_eqn}
	s(t)&=s_0\,\mathrm{rect}\left(\frac{t}{2t_w}\right)\left[b_0+b_1\cos\left(\frac{\pi}{t_w} t\right)+b_2\cos\left(\frac{2\pi}{t_w}t\right)\right]\sin\left(\frac{2\pi k}{t_w}t\right)\\
	&\textnormal{with}\;\mathrm{rect}(t)=\left\{\begin{array}{lr}1\;&\mathrm{for}\;\vert t\vert<\frac{1}{2}\\[\medskipamount]\frac{1}{2}\;&\mathrm{for}\;\vert t\vert=\frac{1}{2}\\[\medskipamount]0\;&\mathrm{for}\;\vert t\vert>\frac{1}{2}\end{array}\right.\textnormal{,}\quad\textnormal{coefficients}\;b_0=\frac{21}{50}\;\textnormal{,}\;b_1=\frac{1}{2}\;\textnormal{,}\;b_2=\frac{2}{25}\;\textnormal{, and}\;k\in \mathbb{N}\;\textnormal{.}\nonumber
\end{align}

The corresponding frequency domain filter function is centered around $f_0=k/t_w$, with maximum $\vert\tilde{s}(f_0)\vert=b_0s_0t_w$ and FWHM $\delta_\mathrm{rbw}\approx 0.822/t_w$ (see Fig.~\ref{Blackman_vs_CPMG} (b)):
\begin{align}
	\tilde{s}(f)=it_ws_0&\left\{b_0\left[\mathrm{sinc}\left(2\pi t_w\left(f+\frac{k}{t_w}\right)\right)-\mathrm{sinc}\left(2\pi t_w\left(f-\frac{k}{t_w}\right)\right)\right]\right.\\
	+&\frac{b_1}{2}\left[\mathrm{sinc}\left(2\pi t_w\left(f+\frac{k+\frac{1}{2}}{t_w}\right)\right)+\mathrm{sinc}\left(2\pi t_w\left(f+\frac{k-\frac{1}{2}}{t_w}\right)\right)\right.\nonumber\\
	&\left.-\mathrm{sinc}\left(2\pi t_w\left(f-\frac{k+\frac{1}{2}}{t_w}\right)\right)-\mathrm{sinc}\left(2\pi t_w\left(f-\frac{k-\frac{1}{2}}{t_w}\right)\right)\right]\nonumber\\
	+&\frac{b_2}{2}\left[\mathrm{sinc}\left(2\pi t_w\left(f+\frac{k+1}{t_w}\right)\right)+\mathrm{sinc}\left(2\pi t_w\left(f+\frac{k-1}{t_w}\right)\right)\right.\nonumber\\
	&\left.\left.-\mathrm{sinc}\left(2\pi t_w\left(f-\frac{k+1}{t_w}\right)\right)-\mathrm{sinc}\left(2\pi t_w\left(f-\frac{k-1}{t_w}\right)\right)\right]\right\}\nonumber\;\textnormal{.}
\end{align}
Note that the suppression of side lobes and harmonics comes at the expense of a reduced peak sensitivity with respect to a uniform square (``CPMG-like'') sequence. The amplification as defined in (\ref{amplification_definition}) is
\begin{equation}
\label{amplification_Blackman}
a_{\tilde{s}}=\left(\int_{-\infty}^\infty\vert \tilde{s}(f)\vert^2df\right)/\delta_\mathrm{rbw}=\frac{t_ws^2_0}{\delta_\mathrm{rbw}}\left(b_0^2+\frac{b_1^2}{2}+\frac{b_2^2}{2}\left(1-\frac{\delta_{k,1}}{2}\right)\right)\approx t_w^2s_0^2\left(0.371- 0.002\delta_{k,1}\right)
\end{equation}
($\delta_{k,1}$ denotes the Kronecker delta in this expression).\\

\subsection{Derivation of the displacement equation of motion (\ref{coherent_state_eom})}
The overall Hamiltonian has the form
\begin{align}
	H(t)&=H_0+H_1(t)+H_2(t)\;\textnormal{, with}\\
	H_0&=\hbar\omega_0a^\dagger a\;\textnormal{,}\\
	H_1(t)&=\hbar\Delta(t) a^\dagger a\;\textnormal{, and}\\
	H_2(t)&=-2\hbar\Omega_d(t)\left(a+a^\dagger\right)\sin(\omega_0 t + \varphi_d)\;\textnormal{.}
\end{align}
We assume $\Delta\ll\omega_0$, and neglect the difference of the eigenstates of $H_0+H_1(t)$ from those of $H_0$.\\

Moving to the interaction picture with respect to $H_0$ (or, equivalently, to the LO reference frame) and neglecting terms rotating at $2\omega_0$, we get
\begin{align}
	H_\mathrm{int}(t)&=H_{\mathrm{int},1}(t) + H_{\mathrm{int},2}(t)\\
	&=\hbar\Delta(t)a^\dagger a + i\hbar\Omega_d(t)\left(a^\dagger e^{-i\varphi_d}-ae^{i\varphi_d}\right)\;\textnormal{,}
\end{align}
with the corresponding propagator
\begin{equation}
U(t_0,t_1)=\exp\left(-i\frac{1}{\hbar}\int_{t_0}^{t_1}H_\mathrm{int}(\tau)d\tau\right)\;\textnormal{.}
\end{equation}
Expressing the integral as a sum over infinitesimal time steps and performing a Trotter decomposition of the exponential \cite{Suzuki1993} (which is exact in the limit $k\rightarrow\infty$) leads to
\begin{align}
	U(t_0,t_1)&=T\left[\exp\left(-i\frac{1}{\hbar}\lim_{k\rightarrow\infty}\sum_{m=1}^k H_\mathrm{int}\left(t_1-m\frac{t_1-t_0}{k}\right)\underbrace{\left(\frac{t_1-t_0}{k}\right)}_{=:\delta t}\right)\right]\\
	&=T\left[\lim_{k\rightarrow\infty}\prod_{m=1}^k\exp\left(-i\frac{1}{\hbar}H_\mathrm{int}(\underbrace{t_1-m\delta t}_{=:t_m})\delta t\right)\right]\\
	&=T\left[\lim_{k\rightarrow\infty}\prod_{m=1}^k\exp\left(-i\frac{1}{\hbar}H_{\mathrm{int},1}(t_m)\delta t\right)\exp\left(-i\frac{1}{\hbar}H_{\mathrm{int},2}(t_m)\delta t\right)\right]\;\textnormal{,}
\end{align}
where the $T$ denotes time ordering of the respective sums and products.\\

Consider the effect of the $m$-th terms on a coherent state $\ket{\alpha}$:
\begin{align}
	\exp\left(-i\frac{1}{\hbar}H_{\mathrm{int},1}(t_m)\delta t\right)\ket{\alpha} &= e^{-\vert\alpha\vert^2/2}\sum_{n=0}^\infty\frac{\alpha^n}{\sqrt{n!}}e^{-i\Delta(t_m)\delta t a^\dagger a}\ket{n}=\ket{\alpha e^{-i\Delta(t_m)\delta t}}\\
	&\approx\ket{\alpha - i\alpha\Delta(t_m)\delta t}\;\textnormal{and}\\
	\exp\left(-i\frac{1}{\hbar}H_{\mathrm{int},2}(t_m)\delta t\right)\ket{\alpha} &= D(\Omega_d(t_m)e^{-i\varphi_d}\delta t)\ket{\alpha}=\ket{\alpha+\Omega_d(t_m)e^{-i\varphi_d}\delta t}\;\textnormal{,}
\end{align}
with the displacement operator $D(\beta)=\exp\left(\beta a^\dagger-\beta^*a\right)$. The equation of motion for $\alpha$ is thus:
\begin{equation}
\label{alphaeom}
\dot{\alpha}=\lim_{\delta t\rightarrow0}\left[\alpha(t+\delta t) - \alpha(t)\right]/\delta t=\Omega_d(t)e^{-i\varphi_d}-i\alpha(t)\Delta(t)\;\textnormal{.}
\end{equation}
In the main text, we set $\varphi_d=0$ for clarity.

\subsection{Analytic expression for the spectral sensitivity}
The formal solution to (\ref{alphaeom}) is \cite{McCormick2019QST}
\begin{equation}
\label{alphaeomsolution}
\alpha(t)=e^{-iI_1(t_0,t)}\left[\alpha(t_0)+\int_{t_0}^t\Omega_d(\tau)e^{-i\varphi_d}e^{iI_1(t_0,\tau)}d\tau\right]\;\textnormal{with}\quad I_1(t_0,t)=\int_{t_0}^{t}\Delta(\tau)d\tau\;\textnormal{.}
\end{equation}
Inserting $\alpha_0(t)$ and $\Omega_d(t)=\dot{\alpha}_0(t)$ into (\ref{alphaeomsolution}) to determine the displacement at $t_w$:
\begin{equation}
\alpha(t_w)=e^{-iI_1(-t_w,t_w)}\left[\underbrace{\alpha(-t_w)}_{=0}+\int_{-t_w}^{t_w}\Omega_d(\tau)e^{-i\varphi_d}e^{iI_1(-t_w,\tau)}d\tau\right]\;\textnormal{.}
\end{equation}
Since a complex phase shift does not affect the observed quantity $\vert\alpha(t_w)\vert^2$, the expression can be simplified as follows:
\begin{align}
	\alpha^\prime(t_w)&=\alpha(t_w)e^{i\left[\pi+I_1(-t_w,t_w)+\varphi_d\right]}\\
	&=-\int_{-t_w}^{t_w}\Omega_d(\tau)e^{iI_1(-t_w,\tau)}d\tau\\
	&=-\underbrace{\left[\alpha_0(\tau)e^{iI_1(-t_w,\tau)}\right]_{-t_w}^{t_w}}_{=0}+\int_{-t_w}^{t_w}\alpha_0(\tau)\Delta(\tau)e^{iI_1(-t_w,\tau)}d\tau\label{int_by_parts}\\
	&\approx\int_{-t_w}^{t_w}\alpha_0(\tau)\Delta(\tau)\left[1+iI_1(-t_w,\tau)\right]d\tau\;\textnormal{.}
\end{align}
The experiments measure the square modulus of this expression, averaged over different noise samples:
\begin{align}
	\left\langle\left\vert\alpha^\prime(t_w)\right\vert^2\right\rangle&=\left\langle\left\vert\int_{-t_w}^{t_w}\alpha_0(\tau)\Delta(\tau)d\tau+i\int_{-t_w}^{t_w}d\tau_2\alpha_0(\tau_2)\Delta(\tau_2)I_1(-t_w,\tau_2)\right\vert^2\right\rangle\\
	&=\left\langle\left\vert\int_{-t_w}^{t_w}d\tau\alpha_0(\tau)\Delta(\tau)\right\vert^2\right\rangle+\left\langle\left\vert\int_{-t_w}^{t_w}d\tau_1\int_{-t_w}^{\tau_1}d\tau_2\alpha_0(\tau_1)\Delta(\tau_1)\Delta(\tau_2)\right\vert^2\right\rangle\;\textnormal{.}\label{coherent_method_sensitivity}
\end{align}
The last step made use of the fact that both integrals are real valued. The analogy of the first term in (\ref{coherent_method_sensitivity}) to (\ref{PSD_result}), 
\begin{equation}
\left\langle\left\vert\alpha^\prime(t_w)\right\vert^2\right\rangle\approx\left\langle\left\vert\int_{-t_w}^{t_w}d\tau\alpha_0(\tau)\Delta(\tau)\right\vert^2\right\rangle\hat{=}\left\langle\left\vert\int_{-t_w}^{t_w}s(t)\Delta(t)dt\right\vert^2\right\rangle=\int_{-\infty}^\infty \left\vert \tilde{s}(f)\right\vert^2S_\Delta(f)df\;\textnormal{,}
\end{equation}
demonstrates the equivalence of $\vert \tilde{\alpha}_0(f)\vert^2$ (where $\tilde{\alpha}_0(f)$ is the Fourier transform of $\alpha_0(t)$) to the spectral sensitivity function $\vert\tilde{s}(f)\vert^2$.\\

The second term of (\ref{coherent_method_sensitivity}) is a correction of order $\Delta^4$. Its response to noise at a frequency $f_n$ can be derived by setting $\Delta(t)=\Delta_0\cos(2\pi f_nt+\varphi_n)$ and averaging over $\varphi_n \in [0,2\pi)$. This corresponds to the response to a PSD of
\begin{equation}
S_\Delta=\frac{\Delta_0^2}{4}\left[\delta(f+f_n)+\delta(f-f_n)\right]\;\textnormal{.}
\end{equation}

\begin{figure}
  \centerline{\includegraphics[width=.8\textwidth]{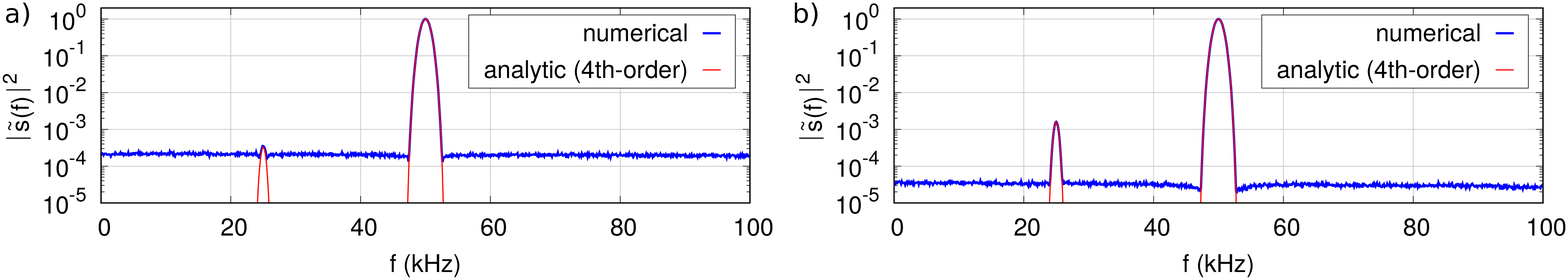}}
  \caption{\label{tickle_sensitivity_supplemental}Numerical and analytic calculation of the spectral sensitivity for coherent displacements along a trajectory as defined by (\ref{blackman_envelope_eqn}), with $t_w=\unit[500]{\mu s}$ and $k=25$. (a) and (b) differ by the noise amplitude used to map out the filter function($\Delta_0=2\pi\times\unit[900]{Hz}$ and $\Delta_0=2\pi\times\unit[2]{kHz}$, respectively), resulting in a more pronounced 4th-order term at $\unit[25]{kHz}$ in (b).}
\end{figure}

For $\alpha_0(t)$ symmetric around $t=0$, the result can be expressed as
\begin{align}
	&\left\langle\left\vert\int_{-t_w}^{t_w}d\tau_1\int_{-t_w}^{\tau_1}d\tau_2\alpha_0(\tau_1)\Delta(\tau_1)\Delta(\tau_2)\right\vert^2\right\rangle=\nonumber\\
	&\frac{\Delta_0^4}{4(2\pi f_n)^2}\left[\left\vert \tilde{\alpha}_0(f_n)\right\vert^2\left(1\pm\frac{1}{2}\cos(2\pi(2f_n)t_w)\right)+\frac{1}{2}\left\vert \tilde{\alpha}_0(2f_n)\right\vert^2\pm \tilde{\alpha}_0(f_n)\tilde{\alpha}_0(2f_n)\cos(2\pi f_nt_w)\right]\;\textnormal{,}
\end{align}
where the upper signs are valid for odd, the lower signs for even symmetry of $\alpha_0$. Figure \ref{tickle_sensitivity_supplemental} shows the agreement of this approximation with numerical calculations for the case of a Blackman-shaped filter function.

\subsection{Experimental implementation}
\subsubsection{Experimental sequence}
The experiments are carried out with a single ${}^9$Be${}^+$ ion confined in a segment of the Paul trap described in \cite{Blakestad2009}. The HO we investigate is the ion motion along the axial direction, with a resonance frequency of $\omega_0=2\pi\times\unit[3.55]{MHz}$. Figure \ref{Qii_term_scheme_square_tickle_supplemental} (a) shows the relevant electronic states and transitions. The sequence is as follows:
\begin{enumerate}
\item State preparation: The motion is first Doppler cooled and then cooled to the ground state via Raman sideband transitions between the $\ket{F=2, m_F=2} =: \ket{\downarrow}$ and $\ket{F=1, m_F=1} =: \ket{\uparrow}$ substates of the ${}^2\mathrm{S}_{1/2}$ ground state manifold. A microwave pulse then transfers the internal state to $\ket{\uparrow}$.
\item Spectrum analyzer sequence: The coherent drive $\Omega_\mathrm{d}$ is applied (see below for details)
\item Detection: To determine the square modulus of the displacement $\vert\alpha(t_w)\vert^2$, a motion-subtracting sideband pulse is applied between $\ket{\uparrow, n}$ and $\ket{\downarrow, n-1}$ with the duration corresponding to a pulse area of $\pi$ on the $\ket{\uparrow,1}\leftrightarrow\ket{\downarrow, 0}$ transition. Finally, the population $P_\downarrow$ in $\ket{\downarrow}$ is detected via resonant light coupling it to a ${}^2\mathrm{P}_{3/2}$ state.
\end{enumerate}

To improve detection fidelity, we insert two microwave ``shelving'' pulse sequences to transfer the population in $\ket{\uparrow}$ to the two substates with the highest energy difference from $\ket{\downarrow}$ before the resonant fluorescence detection pulse: The first transfers population from $\ket{\uparrow}$ to $\ket{F=1, m_F=-1}$ via $\ket{F=2, m_F=0}$, and the second transfers the residual $\ket{\uparrow}$ population to $\ket{F=1, m_F=0}$ (also via $\ket{2, 0}$).

\subsubsection{Observed signals and PSD estimate}
\begin{figure}
  \centerline{\includegraphics[width=1.0\textwidth]{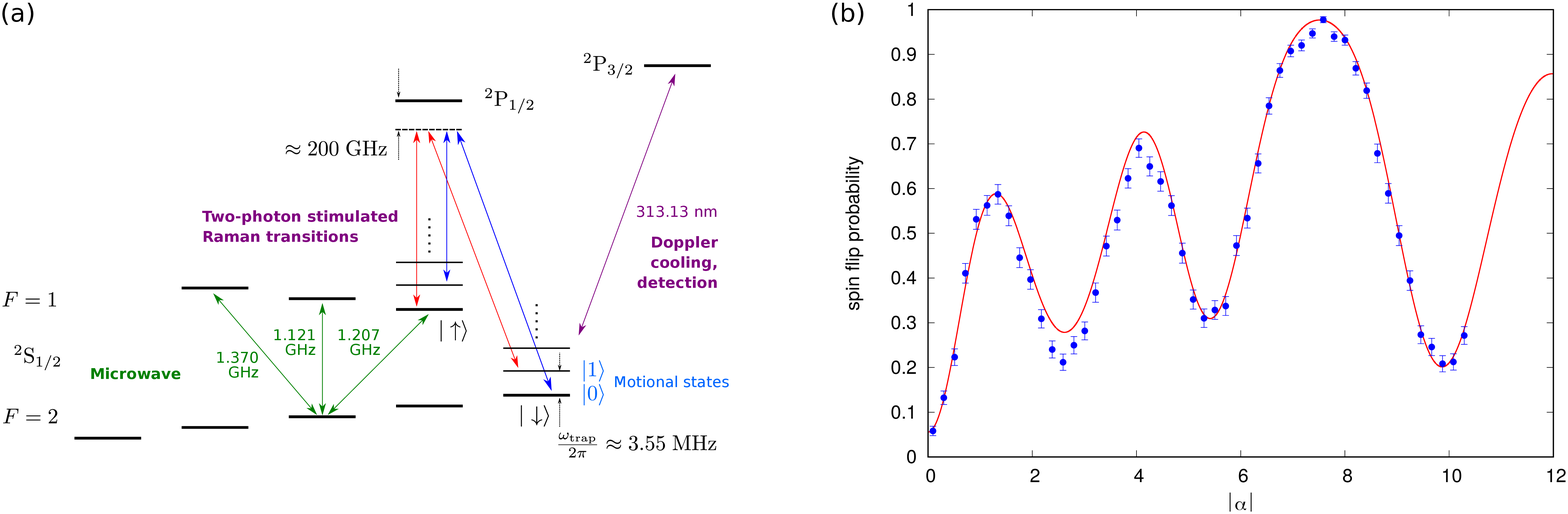}}
  \caption{\label{Qii_term_scheme_square_tickle_supplemental}(a) ${}^9\mathrm{Be}^+$ electronic states used in the coherent state implementation. (b) Spin flip probability when a motion subtracting sideband is applied to a thermal state with $\bar{n}=0.055$ average motional quanta after it has been displaced by $\vert\alpha\vert$. The solid line is the model according to Eq.~(\ref{spin_flip_vs_alpha}). Error bars: $1\sigma$.}
\end{figure}
The detected signal has the form \cite{McCormick2019QST}
\begin{equation}
\label{spin_flip_vs_alpha}
P_\downarrow(\alpha,\bar{n})=\frac{1}{2}\left[1-\sum_{n=0}^\infty P_n(\alpha,\bar{n})\cos\left(\pi\frac{\Omega_{n,n-1}}{\Omega_{1,0}}\right)\right]\;\textnormal{,}
\end{equation}
where $\Omega_{n,m}$ is the Rabi frequency for the transition between motional states $n$ and $m$ \cite{Wineland1998}, and $P_n(\alpha, \bar{n})$ is the population in motional state $n$ for a thermal state of average energy $\bar{n}\hbar\omega_0$, displaced by $\alpha$. An example of such a signal as a function of $\vert\alpha\vert$ for our experimental Lamb-Dicke parameter of $\eta=0.357$ is shown in Fig.~\ref{Qii_term_scheme_square_tickle_supplemental} (b). The finite value of $\bar{n}$ is caused by imperfect ground state cooling and heating throughout the experimental sequence. We determine $\bar{n}$ by replacing the coherent drive with a waiting period of the same duration.\\

The noise power transmitted by the filter corresponds to $\langle\vert\alpha(t_w)\vert^2\rangle$, i.e. the 2nd moment of the distribution of $\vert\alpha\vert$ at the end of the sequence. To avoid the need for assumptions about the shape of this distribution, we require a signal that is unaffected by its higher-order moments. We therefore operate in a regime of $\vert\alpha\vert$ where Eq.~(\ref{spin_flip_vs_alpha}) can be approximated by a 2nd-order polynomial,
\begin{equation}
\label{spin_flip_vs_alpha_poly_approx}
P_{\downarrow}(\vert\alpha\vert, \bar{n})\approx p_0(\bar{n})+p_2(\bar{n})\vert\alpha\vert^2\;\textnormal{.}
\end{equation}
Using (\ref{PSD_approximation}) and (\ref{amplification_Blackman}), we estimate the power spectral density at the filter function center frequency as
\begin{equation}
\label{PSD_estimate_evaluation}
\tilde{S}_\Delta(f_0)=\frac{\langle\phi^2\rangle}{a_{\tilde{s}}\delta_\mathrm{rbw}}\approx\frac{\langle\vert\alpha(t_w)\vert^2\rangle}{0.305s_0^2t_w}\approx\frac{1}{0.305s_0^2t_w}\left(\frac{\left\langle P_{\downarrow}(\vert\alpha(t_w)\vert, \bar{n})\right\rangle-p_0(\bar{n})}{p_2(\bar{n})}\right)\;\textnormal{.}
\end{equation}

\subsubsection{Sensitivity limit}
The smallest detectable displacement $\vert\alpha_\mathrm{min}\vert$ is determined by the uncertainty in measuring $P_\downarrow$ via the requirement
\begin{equation}
P_\downarrow(\vert\alpha\vert)\geq P_\downarrow(\alpha=0) + \sigma_{P_\downarrow}(\alpha=0) \quad\Rightarrow\quad \vert\alpha_\mathrm{min}\vert^2=\frac{\sigma_{P_\downarrow}(P_\downarrow=p_0(\bar{n}))}{p_2(\bar{n})}\;\textnormal{.}
\end{equation}
For the $t_w=\unit[1]{ms}$ sequences used in the measurement shown in Fig.~\ref{coherent_displacement_experiment_figure} (c), $\bar{n}=0.17$, and thus $p_0\approx0.14$, $p_2\approx0.64$. Inserting the shot noise limit for our parameters, $\sigma_{P_\downarrow}(P_\downarrow=p_0)=0.006$, yields $\vert\alpha_\mathrm{min}\vert\approx0.1$. The minimum detectable PSD can be deduced by inserting $\vert\alpha_\mathrm{min}\vert$ into (\ref{PSD_estimate_evaluation}); it scales inversely with the square of the filter function amplitude $s_0^2$. Our limit to the coherent drive voltage results in an $s_0\propto 1/f_0\propto1/k$ dependency (see (\ref{omegamax_eqn}) below). The resolution limit for $t_w=\unit[1]{ms}$ thus becomes
\begin{equation}
\tilde{S}_{\Delta,\mathrm{min}}\approx7.1\times10^{-6}\unitfrac{(rad/s)^2}{Hz}\times k^2=7.1\times10^{-12}\unitfrac{(rad/s)^2}{Hz^3}\times f_0^2\;\textnormal{,}
\end{equation}
as indicated by the shaded region in Fig.~\ref{coherent_displacement_experiment_figure} (c).

\subsubsection{Finite temperature influence and possible quantum enhancement}
\begin{figure}
  \centerline{\includegraphics[width=.8\textwidth]{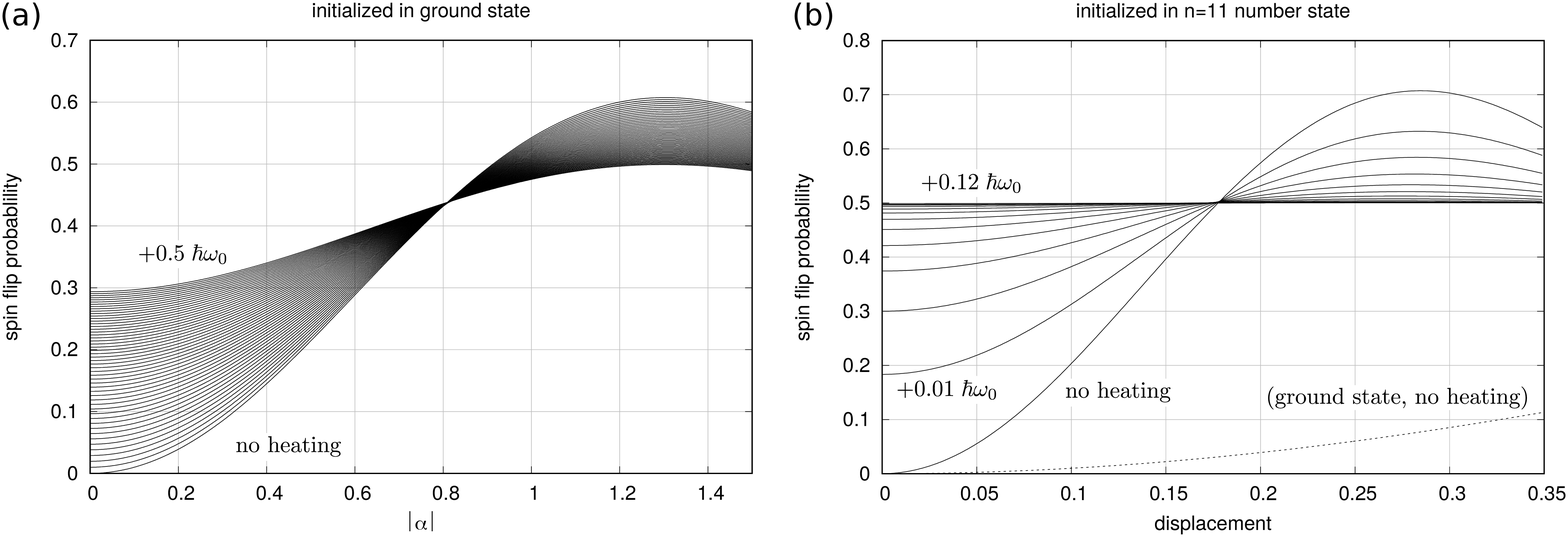}}
  \caption{\label{tickle_heating_Fock_supplemental}Displacement detection signal (Eq.~(\ref{spin_flip_vs_alpha})) under thermal diffusion. (a) Initializing the system in a coherent state. Successive lines show the effect of adding $0.01\;\hbar\omega_0$ of thermal energy to the final state. (b) Possible quantum enhancement of the detection step by initializing the system in a number state (here: $n=11$). The dashed line shows the sensitivity for the ground state (from (a)) for comparison.}
\end{figure}
Figure \ref{tickle_heating_Fock_supplemental} (a) illustrates the effect of finite temperature on the detection signal. Heating corresponds to a diffusion in phase space, which increases the $\ket{n>0}$ population at zero displacement ($p_0$ of Eq.~(\ref{spin_flip_vs_alpha_poly_approx})) and reduces the effect of the displacement itself ($p_2$). Both effects result in a reduced signal-to-noise ratio in the determination of $\vert\alpha\vert^2$.\\

As mentioned in the main text and \cite{McCormick2019thesis,Wolf2019}, the detection can be quantum enhanced by initializing the HO in a pure number state $\ket{n>0}$. This is shown in Fig.~\ref{tickle_heating_Fock_supplemental} (b) for the case of $n=11$ and $\eta=0.3547$. Instead of the first motional sideband, the spin flip probability of the carrier transition is shown, the Rabi frequency of which has a zero crossing at $n=11$. The slope of the signal increases linearly with $n$, allowing the detection of smaller displacements. However, the narrower features of the number state phase space distribution which give rise to this enhancement are more susceptible to the effect of heating, and $0.12$ additional phonons are sufficient to erase the contrast entirely.

\subsubsection{Coherent drive}
We drive the motional state along the trajectory $\alpha_0(t)=s(t)$ as given by equation (\ref{blackman_envelope_eqn}) using a drive of the form $\Omega_d(t)\cos(\omega_\mathrm{LO}t)$, where $\Omega_d(t)=\dot{\alpha}_0(t)$:
\begin{equation}
\Omega_d(t)=s_0\;\mathrm{rect}\left(\frac{t}{2t_w}\right)\sum_{i=1}^5a_i\cos(b_it)\label{blackman_drive_eqn}
\end{equation}
with
\begin{align}
	a&=\frac{2\pi}{t_w} \left(b_0k,\,\frac{b_1}{2}\left(k-\frac{1}{2}\right),\,\frac{b_1}{2}\left(k+\frac{1}{2}\right),\,\frac{b_2}{2}(k-1),\,\frac{b_2}{2}(k+1)\right)^\mathrm{T}\quad\textnormal{and}\label{drive_a_coefficients}\\
	b&=\frac{2\pi}{t_w}\left(k,\,k-\frac{1}{2},\,k+\frac{1}{2},\,k-1,\,k+1\right)^\mathrm{T}\;\textnormal{.}\label{drive_b_coefficients}
\end{align}
The highest displacement in the absence of noise is $s_0$, and the maximum required drive amplitude is
\begin{equation}
\label{omegamax_eqn}
\Omega_{d,\mathrm{max}}=\Omega_d(t=0)=\frac{2\pi k}{t_w} s_0=2\pi f_0 s_0\;\textnormal{.}
\end{equation}

\begin{figure}
  \centerline{\includegraphics[width=.9\textwidth]{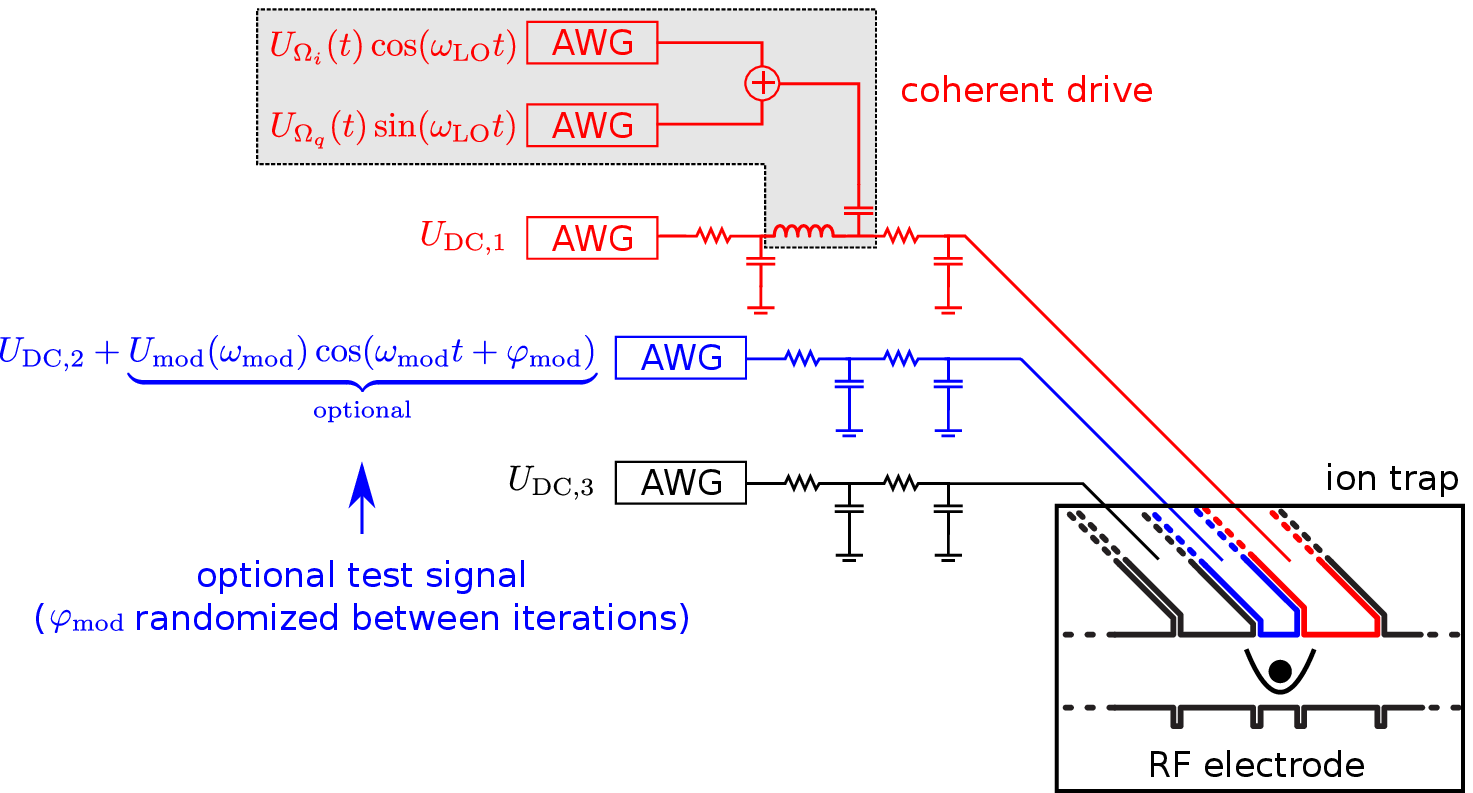}}
  \caption{\label{tickle_setup_supplemental}Electronic setup used for the coherent displacement method. AWG: Arbitrary waveform generator \cite{Bowler2013}. Voltages $U_{\mathrm{DC,1}}$\ldots$U_{\mathrm{DC,3}}$ confine the ion axially (black circle within schematic HO potential). A modulation on $U_{\mathrm{DC,1}}$ implements the coherent displacements via the resulting electric field. The filter functions are characterized with test signals produced by modulating the confinement via a voltage added to $U_{\mathrm{DC,2}}$.}
\end{figure}

Figure \ref{tickle_setup_supplemental} shows the electronic signals applied to the trap electrodes. An electrode neighboring the trapping zone is used to apply the electric fields for the coherent drive, while the electrode in the center of the trapping segment is used for an optional modulation of the HO frequency to diagnose the spectral sensitivity of the method. Lowpass filters suppress electronic noise at the electrodes in order to reduce the heating rate. As the coherent drive waveforms pass these filters, they need to be predistorted, which is achieved by adding two amplitude modulated signals with fundamental frequency $f_0=\omega_\mathrm{LO}/(2\pi)$ in quadrature:
\begin{equation}
U(t)\propto\Omega_i(t)\cos(2\pi f_0t)+\Omega_q(t)\sin(2\pi f_0t)\;\textnormal{.}
\end{equation}

For a 2nd-order RC lowpass, the envelopes have the form
\begin{align}
\Omega_i(t)&=\mathrm{rect}\left(\frac{t}{2t_w}\right)\sum_{i=1}^5 \left[(1-c_2(2\pi f_0)^2)a_i-c_2a_ib_i^2\right]\cos(b_it)-c_1a_ib_i\sin(b_it)\nonumber\\
\Omega_q(t)&=\mathrm{rect}\left(\frac{t}{2t_w}\right)\sum_{i=1}^5 -c_12\pi f_0a_i\cos(b_it)+c_24\pi f_0a_ib_i\sin(b_it)\;\textnormal{, where}\\
c_1&=R_1C_1+(R_1+R_2)C_2\quad\textnormal{,}\nonumber\\
c_2&=R_1C_1R_2C_2\nonumber\;\textnormal{,}
\end{align}
and the $a_i$ and $b_i$ coefficients are those defined in Eqns.~(\ref{drive_a_coefficients}) and (\ref{drive_b_coefficients}). As the first lowpass stage is bypassed in the configuration depicted in Fig.~\ref{tickle_setup_supplemental}, we set $R_1=C_1=0$.

\clearpage
\section{Number state implementation details}
\begin{figure}[h]
	\centerline{\includegraphics[width=1.0\textwidth]{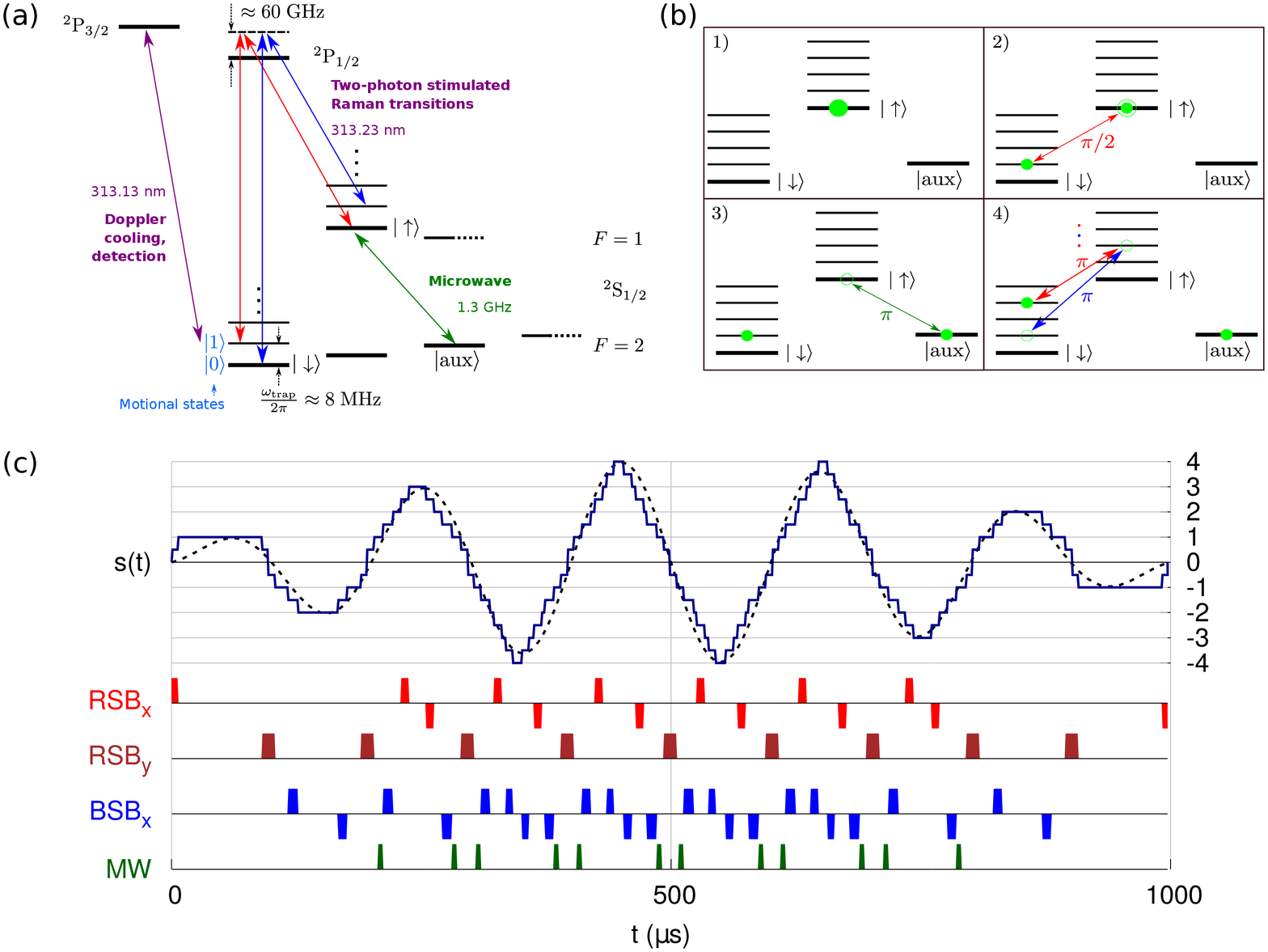}}
	\caption{\label{Fock_sequence_supplemental}Pulse sequences for the number state method. (a) ${}^9\mathrm{Be}^+$ electronic states involved in the pulse sequences. (b) Principle of a motional $\pi/2$ pulse. (c) Spectrum analyzer sequence for the $\unit[5]{kHz}$ filter function. The dashed line shows the target time-domain sensitivity function. An optimization algorithm generates a sequence of pulses, shown below, to approximate it with a piecewise constant function (solid line). RSB: Red sideband Raman transition, BSB: Blue sideband Raman transition, MW: Microwave transition.}
\end{figure}
The number state experiments are carried out in the surface electrode Paul trap described in \cite{Brown2011}. Figure \ref{Fock_sequence_supplemental} (a) shows the involved electronic states and transitions: $\ket{F=2, m_F=-2} =:\ket{\downarrow}$, $\ket{F=1, m_F=-1} =:\ket{\uparrow}$, and $\ket{F=2, m_F=0} =:\ket{\mathrm{aux}}$. Transitions between $\ket{\downarrow, n}$ and $\ket{\uparrow, n\pm1}$ are implemented as Raman transitions coupling off-resonantly to the ${}^2\mathrm{P}_{1/2}$,${}^2\mathrm{P}_{3/2}$ states, and microwave pulses are used for coupling $\ket{\uparrow,n}\leftrightarrow\ket{\mathrm{aux},n}$. State detection is achieved by resonantly coupling $\ket{\downarrow}$ to ${}^2P_{3/2}$.\\

An example sequence to generate a motional superposition is shown in Fig.~\ref{Fock_sequence_supplemental} (b): After preparation in $\ket{\uparrow,0}$ by Raman ground state cooling (step 1), a $\frac{\pi}{2}$-pulse on the $\ket{\uparrow, 0}\leftrightarrow\ket{\downarrow,1}$ motional sideband splits the population (step 2). After ``shelving'' the $\ket{\uparrow}$ population to $\ket{\mathrm{aux}}$ (step 3), further motional sideband $\pi$-pulses move the other superposition component to successively higher number states (step 4). For target states up to $n=2$, the shelving step can be omitted.\\

Figure \ref{Fock_sequence_supplemental} (c) shows the pulse sequence used to generate the $\unit[5]{kHz}$ filter function from the main text. Delays are added between the pulses in step 4 to approximate the target filter shape. The pulse phases are chosen to improve robustness against pulse imperfections. In each individual lobe, the phase of a ``downward'' sideband pulse is shifted by $\pi$ with respect to the corresponding ``upward'' pulse (as indicated by the different sign). The $\pi$-pulses between different lobes are phase shifted by $\pi/2$ with respect to the separation and recombination $\frac{\pi}{2}$-pulses at the beginning and end of the sequence (denoted by the indices $x$ and $y$), analogous to the phase shifts used between the $\frac{\pi}{2}$ and $\pi$ pulses of CPMG sequences.\\

The sensitivity function of such a sequence can be approximated as follows: While the HO is in a superposition of $\ket{n_1}$ and $\ket{n_2}$, $s(t)$ takes the value $n_2-n_1$. For a function $s(t)$ that takes the value $s(t)=s_i$ for $t_i\leq t<t_{i+1}$ with $i=1\ldots N$, the frequency domain filter function becomes
\begin{equation}
\label{fock_method_spectral_sensitivity}
\vert \tilde{s}(f)\vert^2=\frac{1}{(2\pi f)^2}\sum_{i,j=1}^{N+1}\left(s_{i-1}-s_i\right)\left(s_{j-1}-s_j\right)e^{i2\pi f(t_i-t_j)}\quad\textnormal{with}\quad s_0=s_{N+1}=0\;\textnormal{.}
\end{equation}

For filter frequencies approaching the limit set by the minimum pulse durations, phase accumulation due to $\Delta(t)$ during pulse application needs to be taken into account. The simplest assumption is for $s(t)$ to be constant at a value halfway between those before and after the pulse, neglecting the pulse area error due to the detuning. The analytic sensitivity function shown in Fig.~\ref{Fock_filter_theory} (b) is calculated using (\ref{fock_method_spectral_sensitivity}) under this assumption. The numerical trace in the same graph takes the full effect of detuned pulses into account.\\

Experimental pulse sequences are generated by an optimization algorithm that adjusts pulse sequence parameters within experimental constraints to match such a piecewise constant sensitivity function to a smooth target function (shown in Figs.~\ref{Fock_filter_theory} (a) and \ref{Fock_sequence_supplemental} (c) as a dashed line).\\

 The impact of pulse imperfections on noise sensitivity increases with the fraction of sequence duration spent within pulses. As the microwave parameters are very stable, we are mainly concerned with a global mismatch in Rabi frequency of the optical transitions, e.g.~from beam pointing fluctuations. Numerical calculations applying white noise around the filter passband yield a sensitivity of $\Delta P_\mathrm{bright}/P_\mathrm{bright}\approx 5 \Delta\Omega_\mathrm{Rabi}/\Omega_\mathrm{Rabi}$ for the $\unit[5]{kHz}$ sequence, which consists almost entirely of pulses. For the $\unit[500]{Hz}$ sequence, this sensitivity reduces to $\Delta P_\mathrm{bright}/P_\mathrm{bright}\approx 0.7 \Delta\Omega_\mathrm{Rabi}/\Omega_\mathrm{Rabi}$ . In the experiments, these errors are mitigated by interleaved Rabi frequency calibrations.

\end{document}